\documentclass[11pt]{article}
\usepackage[utf8]{inputenc}

\usepackage{fancyhdr,amsmath,amsthm,amssymb,bm,url,enumerate,float,subfigure,multicol, multirow} 
\usepackage{lastpage} 
\usepackage{extramarks} 
\usepackage{graphicx,caption} 
\usepackage{microtype}
\usepackage{listings}
\usepackage{cancel}
\usepackage{graphicx} 
\usepackage{caption}
\usepackage{geometry}
\usepackage{listings}
\usepackage{xcolor}
\usepackage{natbib}
\usepackage{authblk}
\usepackage[nodisplayskipstretch]{setspace}
\usepackage[ruled,vlined,linesnumbered]{algorithm2e}
\usepackage[]{hyperref}
\hypersetup{
    pdftitle={Your title here},
    pdfauthor={Your name here},
    pdfsubject={Your subject here},
    pdfkeywords={keyword1, keyword2},
    bookmarksnumbered=true,     
    bookmarksopen=true,         
    bookmarksopenlevel=1,       
    colorlinks=true, 
    citecolor=blue,
    urlcolor=blue,
    pdfstartview=Fit,           
    pdfpagemode=UseOutlines,    
    pdfpagelayout=TwoPageRight
}

\newcommand{\Title}{Difference-in-Differences under Bipartite Network Interference:\\ A Framework for Quasi-Experimental Assessment of the Effects of Environmental Policies on Health}

\newtheorem{assumption}{Assumption}
\newtheorem{proposition}{Proposition}

\newcommand{\EE}{\mathbb{E}}

\newcommand{\II}{\mathbb{I}}

\def\<#1>{\mathinner{\langle#1\rangle}}

\title{\Title}
\author[1]{Kevin L. Chen}
\author[1]{Falco J. Bargagli-Stoffi}
\author[1]{Raphael C. Kim}
\author[2]{Lucas R.F. Henneman}
\author[1]{Rachel C. Nethery}
\affil[1]{Department of Biostatistics, Harvard T.H. Chan School of Public Health (Boston, MA)}
\affil[2]{Department of Civil, Environmental, and Infrastructure Engineering, George Mason University (Fairfax, VA)}
\date{}

\begin{document}

\maketitle

\begin{abstract}
Pollution from coal-fired power plants has been linked to substantial health and mortality burdens in the US. In recent decades, federal regulatory policies have spurred efforts to curb emissions through various actions, such as the installation of emissions control technologies on power plants. However, assessing the health impacts of these measures, particularly over longer periods of time, is complicated by several factors. First, the units that potentially receive the intervention (power plants) are disjoint from those on which outcomes are measured (communities), and second, pollution emitted from power plants disperses and affects geographically far-reaching areas. This creates a methodological challenge known as bipartite network interference (BNI). To our knowledge, no methods have been developed for conducting quasi-experimental studies with panel data in the BNI setting. In this study, motivated by the need for robust estimates of the total health impacts of power plant emissions control technologies in recent decades, we introduce a novel causal inference framework for difference-in-differences analysis under BNI with staggered treatment adoption. We explain the unique methodological challenges that arise in this setting and propose a solution via a data reconfiguration and mapping strategy. The proposed approach is advantageous because analysis is conducted at the intervention unit level, avoiding the need to arbitrarily define treatment status at the outcome unit level, but it permits interpretation of results at the more policy-relevant outcome unit level. Using this interference-aware approach, we investigate the impacts of installation of flue gas desulfurization scrubbers on coal-fired power plants on coronary heart disease hospitalizations among older Americans over the period 2003-2014, finding an overall beneficial effect in mitigating such disease outcomes.
\end{abstract}


\pagebreak


\newpage
\section{Introduction}\label{sec:intro}

\doublespacing
Coal-fired power plants are significant emitters of pollutants, including sulfur dioxide (SO$_2$) and nitrous oxides (NO$_x$), which are also key contributors to the formation of fine particulate matter smaller than $2.5 \mu$g in size (PM$_{2.5}$). Exposure to power plant-related PM$_{2.5}$ has been linked to an increased risk of cardiovascular and respiratory diseases, as well as stroke, and has been estimated to be responsible for 460,000 deaths between 1999 and 2020 among U.S. Medicare beneficiaries \citep{Samet_mortality_2000,Tsai_stroke_2003,Koken_cvd_2003,Dominici_cvdrd_2006,Wu_mortality_2020,Henneman_mortality_2023}. PM$_{2.5}$ exposure has also been shown to be a risk factor for cancer, neurodegenerative diseases, and pregnancy complications, among other adverse health outcomes \citep{Dadvand_preg_2013,Hamra_lung_2014,Feng_effects_2016,Shou_alz_2019,Xie_still_2021,Cristaldi_neuro_2022}. 

As a direct result of this growing body of research demonstrating the detrimental health effects of both short-term and long-term air pollution exposure, regulatory authorities such as the U.S. Environmental Protection Agency (EPA) have enacted policies aimed at curbing the amount of ambient air pollution to protect public health and welfare. The 1990 Amendments to the Clean Air Act and subsequent changes to the National Ambient Air Quality Standards (NAAQS) set limits on ambient levels of six pollutants, including SO$_2$, NO$_x$, and PM$_{2.5}$. Localities that do not meet these standards, or ``nonattainment areas'', may be assessed penalties under the NAAQS rules. These policy enactments have led to increasing measures being taken to reduce power plant emissions in order to remain in compliance with regulations, such as switching to alternative fuel types or the employment of emissions control technologies \citep{Raff_naaqs_2019}.

This paper is motivated by our aim of robustly quantifying the health impacts emissions-reducing interventions on power plants in order to inform policy evaluation and future policy development. In general, the challenge with assessing the impact of power plant emissions on human health has been the complex mechanisms and pathways through which emitted pollutants are transported. Primary PM$_{2.5}$ that is directly emitted from power plants can disperse through wind and linger airborne for long periods of time, while secondary PM$_{2.5}$ is additionally formed in atmospheric chemical conversions of other emitted gases \citep{Wilson_pm_1997}. Consequentially, emissions from a power plant have an effect on pollution concentrations not only in its immediate surrounding area, but potentially also in communities hundreds or thousands of miles away. The result is a complex web of associations that can be depicted as a dense network of connections between power plant emission sources and human populations. Specifically, this is what is known as a bipartite network---a network that consists of two disjoint sets of units (or ``nodes''), where connections (or ``edges'') are allowed strictly between the sets, but not within a set. Such networks are common in the social sciences---for example, one representation of a co-stardom network consists of two groups: actors and movies. Edges in this network exist if an actor stars in a movie but not otherwise.

In our motivating application, we have two distinct sets of units: power plants on which interventions (or ``treatments'') are potentially imposed and communities in which health outcomes are measured. We will refer to these two sets of units more generally as ``intervention units'' and ``outcome units'', respectively. The units are connected in a bipartite network, i.e., each power plant is connected to many (potentially distant) communities based on the pollution transport from the power plant to the community, but there are no power plant-to-power plant or community-to-community connections. This data structure dramatically complicates policy evaluation. Note that in the context of power generation, the fundamental level of electric generating unit (EGU) data is the ``unit''; however, in this paper we refer to the ``unit'' from the statistical perspective--- that is, an individual observation from the data. 

In this bipartite network, an intervention at one power plant may affect many downwind communities and each community may be affected by the intervention status of multiple power plants. To evaluate the effects of such an intervention in a causal inference framework, one inevitably confronts the fundamental challenge known as ``treatment interference'' or ``spillover effects'', i.e., the setting in which an intervention on one unit may impact the outcomes of other units \cite[seminar reference to this phenomenon can be found in][]{cox1958planning}. The presence of interference violates the often held causal assumption known as the Stable Unit Treatment Value Assumption (SUTVA), which requires that the outcomes are only affected by the unit's own treatment \citep{Rubin_sutva_1986}. Causal analyses in which outcomes depend on the treatments of other units require attention to interference to fully account for their effects, and otherwise could lead to misleading inference and results \citep{Sobel_interference_2006,Karwa_interference_2018,Forastiere_interference_2020, Bargagli_clusterednetwork_2020}. 

Causal inference in a setting that combines the concepts discussed above---that is, the presence of treatment interference defined by a bipartite network---was first discussed in \cite{Zigler_bipartite_2021}. In the remainder of this paper, we refer to this general setting as that of \textit{bipartite network interference} (BNI), as in \cite{Chen_bipartite}. Although several previous papers have developed methods for conducting causal inference in cross-sectional observational study designs with BNI, to our knowledge, none have considered how to address BNI in quasi-experimental study design settings accounting for different intervention times (also known as ``staggered treatment adoption''). In this paper, motivated by a study of the health impacts of power plant emissions reduction interventions implemented over the course of a decade in the U.S., we propose an approach for performing causal difference-in-differences (DiD) analysis with panel data under BNI.

\subsection{Motivating Application}\label{sec:motivating}

Flue gas desulfurization (FGD) equipment, also known as ``scrubbers'', have been instrumental tools in lowering the emissions of power plants in the U.S. since the 1970s, with reported SO$_2$ removal efficiencies of up to 99\% \citep{sorrels_scrubbers}. Reductions in SO$_2$ emissions have subsequently been found to result in substantial benefits to cardiac and respiratory health outcomes \citep{Henneman_health_2019}. Several recent studies have examined the causal effect of scrubber installations on various human health outcomes at particular points in time from a BNI perspective, but to date no work has examined these effects comprehensively over time \cite[see, e.g.,][]{Zigler_bipartitetransport,Chen_bipartite}. Since each power plant and locality independently determine the appropriate measures for emissions control, the timing of scrubber installations has varied over the past few decades. Another factor to consider is the significant cost of retrofitting this technology onto existing power plants. Although exact costs vary depending on the location, space limitations, or operating conditions of a power plant, initial capital costs have been estimated to often exceed hundreds of millions of dollars, with additional tens of millions of dollars in annual maintenance and labor costs \citep{usepa_scrubberfactsheet_2003,sorrels_scrubbers}. 

It is, therefore, of considerable importance for regulatory and policymakers to understand the overall impacts of the installation of emissions control technology to inform future development of cost-effective policies to protect health. To more rigorously and comprehensively characterize the effect of FGD scrubber installation on power plants in the years since NAAQS implementation, we use a quasi-experimental DiD design to estimate the effect of FGD scrubber installation on coronary heart disease (CHD) hospitalizations in older Americans in the contiguous U.S. between 2003 and 2014. The use of quasi-experimental designs when treatment randomization is not possible (that is, in most health policy applications) provides stronger protections against potential unmeasured confounding by accounting for certain types of confounding by design \citep{Rockers_quasi_2015,Wing_quasi_2018}. Studies using these more robust approaches often receive higher weight in informing policymaking considerations and, thus, are of substantial interest when evaluating past large-scale policy decisions.

The health outcome data used in this study are annual  CHD hospitalization rates among Medicare Fee-for-Service beneficiaries residing in 39,474 U.S. ZIP codes over the period 2003-2014. These were obtained from the Medicare Provider Analysis and Review files from the U.S. Center for Medicare and Medicaid Services. Information on power plant facility operation and scrubber installation for 383 coal-fired power plants continuously operating during the study period was obtained from the U.S. EPA Air Markets Program Database. We characterize the BNI structure using a reduced complexity pollution transport model called the Hybrid Single-Particle Lagrangian Integrated Trajectory (HYSPLIT) Average Dispersion model \citep{Henneman_hyads_2019}, or HyADS for short. HyADS delineates the level of exposure of each ZIP code to emissions from each individual coal-fired power plant, which is represented by a source-receptor matrix. For this work, we use a modified version of the HyADS matrix in which the edge weights of the interference network are not affected by emissions control strategies at power plants (the importance of this feature for our methodological framework is described in more detail below) that is used as the bipartite interference network in this study. The entries in this modified HyADS matrix can be interpreted as the levels of power plant emissions exposure on each ZIP code that is influenced by the atmospheric and meterological conditions which drive the transport of pollution from emissions sources.

The remainder of the paper is organized as follows. Section~\ref{sec:did_existing} describes the existing difference-in-differences methodology and extensions of this work to more complex scenarios. Section~\ref{sec:did_bni} proposes a new framework to adapt the data and existing methodology to estimate causal effects using the DiD methodology in bipartite interference settings. Section~\ref{sec:case} presents an application of this proposed framework to a case study based on the motivation application described above.  Section \ref{sec:conclusion} concludes the paper.

\section{Difference-in-Differences: Existing Approaches and Considerations}\label{sec:did_existing}

In this section, we provide a brief overview of DiD designs and existing methodology, and then discuss some challenges and considerations when exploring extensions of these methods to the BNI setting. Several recent review papers \citep{Freedman2023,Roth2023} detail the various methodologies and recent advances in DiD, and we refer the reader to these works for a more comprehensive review of the existing literature.

\subsection{Canonical DiD}\label{sec:canonical}

We begin with the so-called ``canonical'' DiD setting, with two time periods, $t \in \{1,2\}$, and two treatment groups. All units, $j=1,...,N$, are untreated at time $t=1$ and a subset of them becomes treated between times 1 and 2. Define $D_j$ as an indicator of whether a unit is ever treated, that is, $D_j = 1$ for those units that are treated at time $t=2$, and $D_j=0$ for those never treated. The outcomes are observed for all units at both time points and are denoted by $Y_{jt}$. Under the potential outcomes framework \citep{Rubin1974}, the average treatment effect on the treated (ATT) is defined as 
\[\tau = \EE[Y_{j2}(1) - Y_{j2}(0) \mid D_j = 1],\] 
where $Y_{jt}(1)$ is the potential outcome for unit $j$ at time $t$ when the unit is treated at time $t=2$, while $Y_{jt}(0)$ is the potential outcome for unit $j$ at time $t$ when the unit is not treated at time $t=2$. In reality, this design means that untreated potential outcomes are observed for all units at time $t=1$, while at time $t=2$, $Y_{j2}(1)$ is observed for treated units and $Y_{j2}(0)$ is observed for untreated units. Thus, the fundamental problem for the identification of the ATT is that $Y_{i2}(0)$ is not observable for treated units. DiD methods resolve this by utilizing a \textit{parallel trends assumption}, which assumes that in the complete absence of treatment, the expected outcomes of treated and untreated groups would have changed in parallel over time. Under parallel trends, combined with the \textit{no anticipation assumption}, the identification of $\EE[Y_{j2}(0) \mid D_j=1]$ can easily be shown. In practice, estimation is often performed using a two-way fixed effects (TWFE) regression: 
\[Y_{jt} = \mu_j + \lambda_t + D_j \cdot \II(t=2)\beta + \varepsilon_{jt}\]
in which $\mu_j$ is a unit fixed effect, and $\lambda_t$ is a time fixed effect. Ordinary least squares can be used for estimation of this model, and the resulting $\widehat{\beta}$ is a consistent estimate of $\tau$.

However, for many real-world applications, the limitation in the canonical setting of just two time periods and two treatment groups is too restrictive to obtain any meaningful insights. Consider our motivating application as an example: It would be unreasonable to assume that there are only two time periods and that all treated power plants were treated in the same period. Scrubber installations have been occurring for decades, with different power plants having installations done in different years. Furthermore, the effects of scrubber installation are likely to vary depending on factors such as the time of installation and/or how long the scrubbers have been present. It is therefore of interest to consider methodology designed for multi-period settings with \textit{staggered} treatment adoption.

\subsection{Generalizations of Canonical DiD}\label{sec:staggered}

The TWFE specification in Section~\ref{sec:canonical} can be extended to the multi-period, staggered treatment adoption setting in a straightforward manner. Now let $t \in \{1,\dots,T\}$ index the time period and $D_{jt}$ denote an indicator for whether the unit $j$ is treated at time $t$, with $F_j$ being the time period in which the unit is first treated. In this scenario, let $Y_{jt}(f)$ be the potential outcome for unit $j$ at time $t$ when the unit was first treated at time $f$ and let $Y_{jt}(\infty)$ be the potential outcome for a unit that was never treated during the study period. In the most general setting, we can define the ATT at time $t$ for a ``cohort'' of units first treated at time $f$ as 
\[ \tau (f,t) = \EE[Y_{jt}(f) - Y_{jt}(\infty) \mid D_{jt} = 1, F_j=f].\]
This general estimand was first proposed by \cite{Callaway2021} and \cite{Sun2021}, and it permits maximal flexibility by allowing for heterogeneity across cohorts and time. Several variants of it, obtained by collapsing across cohorts and/or time, have also been considered and estimated using extensions of TWFE.

If we are willing to assume that $\tau (f,t)=\tau$ for all $(f,t)$, i.e., that the ATT is constant across time and cohorts, the following ``static'' TWFE model can be used for estimation: \[Y_{jt} = \mu_j + \lambda_t + D_{jt}\beta + \varepsilon_{jt}.\] When treatment effects are indeed homogeneous across cohorts and time, this specification produces a $\widehat{\beta}$ that is a consistent estimator for $\tau$. However, this scenario is overly simplistic and unrealistic in most cases, and the recent literature has shown that the TWFE approach does not result in interpretable effects of interest when heterogeneity does in fact exist \cite[see, e.g.,][]{GoodmanBacon2021, Sun2021, deChaisemartin2022, borusyak_revisiting_2023}.

Therefore, we may consider alternative methods for more flexible estimation of treatment effects. One such extension allows for heterogeneity across time relative to treatment initiation, and this ``dynamic'' TWFE specification can be written as follows: 
\[Y_{jt} = \mu_j + \lambda_t + \sum_{k = -L}^{M} D_{jt}^k\beta^k + \varepsilon_{jt},\]
where $k \in \{-L,\dots,-1\}$ indicate lead times prior to treatment initiation, $k=0$ indicates the period in which treatment begins, $k \in \{1,\dots,M\}$ indicate lag times post-treatment, and $D_{jt}^k$ is an indicator of relative time since treatment. The resulting $\beta^k$ is an estimate of the average treatment effect after exposure to $k$ periods of treatment. This is a variant of $\tau (f,t)$ collapsed across $(f,t)$ representing a common time since treatment initiation. As with static TWFE, however, similar issues with this particular dynamic TWFE specification have been shown when effects after $k$ periods of treatment differ depending on when treatment was initiated \citep{Sun2021}.

To overcome this limitation, \cite{Callaway2021} (using inverse proability weighting and doubly robust approaches) and \cite{Sun2021} (using a TWFE approach) proposed to estimate the general $\tau (f,t)$, allowing for heterogeneity across both cohorts and time. The various $\tau (f,t)$s can then be aggregated to estimate some average effect parameter of interest, such as the collapsed estimands described above, if desired. \cite{Gormley_stacked_2011}, \cite{Deshpande_stacked_2019}, and \cite{Cengiz_stacked_2019} implement a stacked regression approach, which \cite{Gardner_twoway} shows estimates a weighted average of $\tau (f,t)$ terms. As another alternative, \cite{borusyak_revisiting_2023} and \cite{Gardner_twoway}, among others, propose instead ``imputation''-type estimators, in which a TWFE regression model is fit for control units (either never or not-yet-treated observations). This regression can then be used to impute the counterfactual ``untreated'' outcomes for treated units and aggregated, or, in the case of \cite{Gardner_twoway}, regressed on treatment indicator values to obtain ATT estimates.

Several works in the literature have also proposed methodology for DiD analyses accounting for interference and allowing for characterization of spillover effects \citep{VerbitskySavitz_spatial,Delgado_spatial,Clarke2017,Bergspatial}, while \cite{Butts2021} further extends the literature to staggered adoption settings. We focus here on the latter, particularly for event study specifications. \cite{Butts2021} defines potential outcomes as $Y_{jt}(D_{jt},g_j(\overrightarrow{D_t}))$, where $D_{jt}$ is a treatment indicator for whether unit $j$ is treated, and $g_j(\overrightarrow{D_t})$ is some summary function of the vector of all units' treatment assignments $\overrightarrow{D_t} \in \{0,1\}^n$, using the ``exposure mapping'' idea of \cite{Aronow2017}. This formulation formalizes the concept that, in the presence of interference, the relevant ``treatment'' for each unit consists of both the unit's own individual treatment assignment and a function mapping of other units' treatment assignments that encapsulates the spillover experienced by the given unit. The result, depending on the exact mapping strategy, allows for the identification of control units in a DiD analysis, i.e., units for which both $D_{jt}=0$ and $g_j(\overrightarrow{D_t})=0$. This allows for estimation of ATT-like quantities using imputation estimators, following \cite{Gardner_twoway} and \cite{borusyak_revisiting_2023}. Our proposed estimands and estimation approaches for DiD under BNI follow from this strain of work, and are further detailed in Section~\ref{sec:did_bni}.

\subsection{Bipartite Network Interference Considerations}\label{sec:bni_considerations}

The nature of bipartite exposure-response networks adds yet another layer of complexity for DiD-type analyses of this nature. The most apparent challenge when looking at such a problem is how to reconcile two distinct sets of intervention and outcome units. In this scenario, when applying the existing DiD methodology, two possible perspectives may be considered. The first focuses on the \textit{outcome unit}, which is perhaps the most intuitive of the two if one's goal is to estimate changes in the outcome. The second focuses on the \textit{intervention unit}, which we discuss in detail in Section~\ref{sec:did_bni}. The outcome unit perspective is the approach taken in existing cross-sectional bipartite interference studies and carries the advantage of being able to work with measured outcomes directly (e.g., health outcomes on the ZIP code-level). However, this introduces a new consideration: what does it mean for a unit to be ``treated''? In the context of our motivating problem, estimates of health effects would be measured at the ZIP code-level, but each ZIP code would have multiple associated treatments (i.e., an individual treatment value for each power plant to which it is connected). 

In previous cross-sectional work in the BNI context, \cite{Zigler_bipartitetransport} and \cite{Chen_bipartite} map the associated treatment vector for each outcome unit to a direct and indirect component (referred to as ``key-associated'' and ``upwind'' in an air pollution exposure context). \cite{Doudchenko_bipartite} and \cite{Harshaw_bipartite} propose instead mapping the intervention unit treatment vector to a continuous-valued exposure for each outcome unit. The application of these approaches, however, is not straightforward to extend to a panel data setting. DiD-type methods usually rely on ``control'' units serving as comparison groups for units that were treated. Under these mapping approaches in the bipartite setting, it becomes apparent that the idea of ``treated'' and ``untreated'' becomes difficult to define. For example, the $\tau (f,t)$ estimand proposed by \cite{Callaway2021} captures the average effect of treatment at a given time point for cohorts of treatment units (that is, groups of units that were first treated in the same period). It is not readily apparent, however, how this concept would be applied to the bipartite setting. Furthermore, this introduces a potential disconnect with relevant future interventions---treatment effects are measured on the outcome units, but from a policy perspective, treatments can only be imposed on intervention units. For instance, environmental policymakers may be interested in effects on cardiovascular-related disease hospitalizations but only have the ability to intervene on emitting power plants.

The alternative is to take the perspective of the intervention unit instead. This perspective has two main advantages. First, when working with panel data in a DiD-type analysis, this has the distinct advantage of retaining the notion of each unit having its ``own'' treatment. This also allows us to more easily leverage the vast swath of existing DiD methodology which has been designed to estimate the effects of a unit's direct treatment. However, this approach necessitates a projection of the information observed on the outcome units onto the intervention units, for which we detail one potential method in Section~\ref{sec:did_bni}. Second, as highlighted above, this approach has a clear advantage for policy-making in that it allows one to take the perspective of the same units on which the policy-makers can effectively deploy their interventions. As a bonus, we show that the proposed estimands at the intervention unit level can also be interpreted as public health-relevant estimands at the outcome unit level.

 \section{A Difference-in-Differences Framework under BNI}\label{sec:did_bni}

In this section, we propose a novel and general framework for reconfiguring BNI-structured data to enable application of existing methods for interference-aware DiD-type quasi-experimental analysis, at the intervention unit level. For interpretability and to demonstrate the applicability of this framework, we detail the approach in the context of our motivating application and, henceforth, refer to the intervention units as power plants, to outcome units as ZIP codes, and the edge weights of the interference network as ``HyADS values''. As preliminaries, we first define the notation and detail a mapping procedure for intervention-level outcomes. Then, we outline a number of assumptions, as well as strategies for estimation and identification of comparison control units. 

\begin{figure}[ht!]
    \centering
    \includegraphics[width=\textwidth]{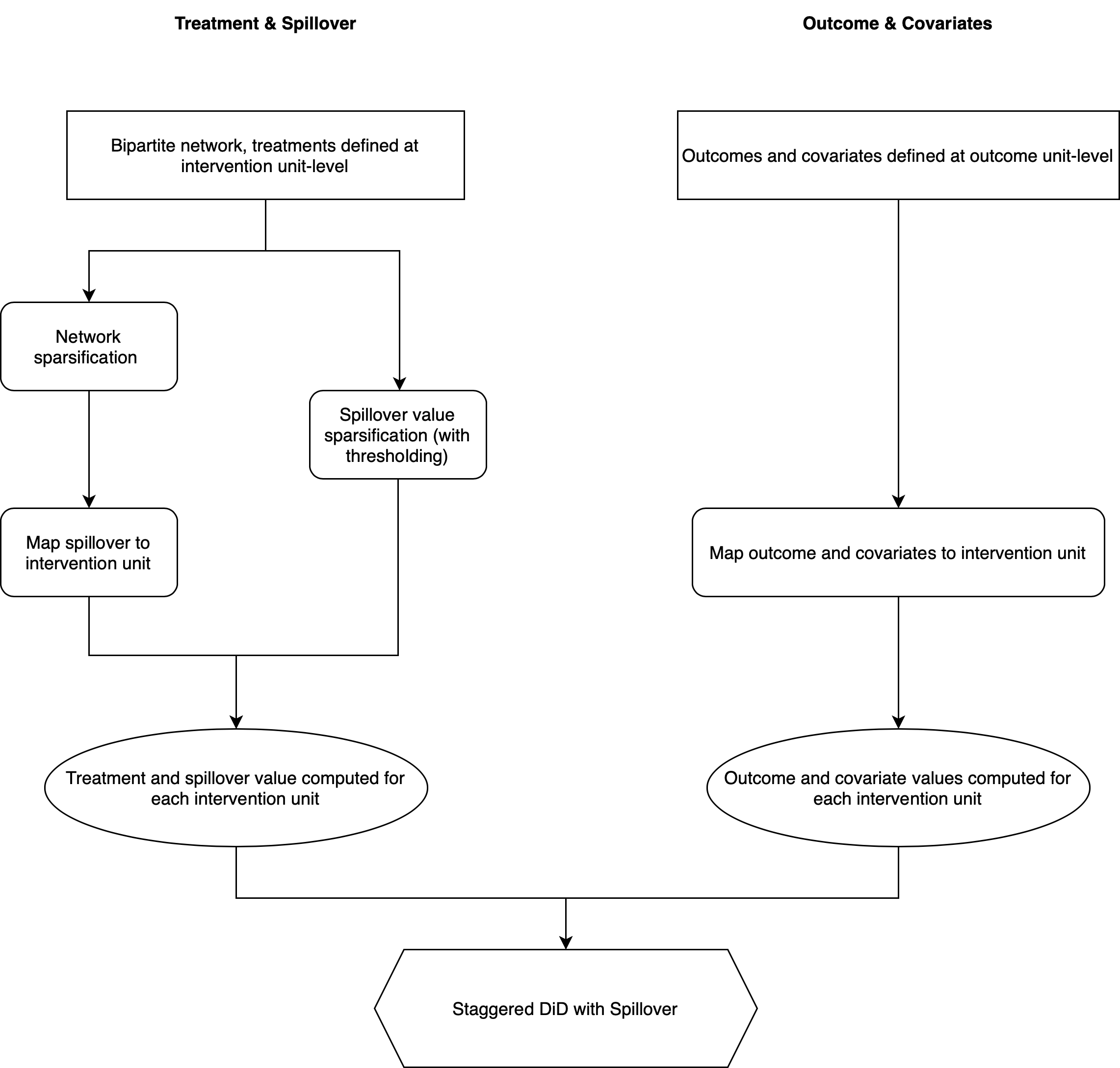}
    \caption{Overview of proposed framework.}
    \label{fig:overview}
\end{figure}

This proposed framework is displayed graphically in Figure~\ref{fig:overview}.

\subsection{Notation and Outcomes}\label{sec:notation}

Let $j \in \{1,\dots,J\}$ index the set of power plants, $i \in \{1,\dots,N\}$ index the set of ZIP codes, and $t \in \{1,\dots,T\}$ denote the time period. Let $\bm{A}\in \{0,1\}^{J \times T}$ be the matrix of all treatment statuses of the power plant over time, with the element $a_{jt}$ being treatment status of the power plant $j$ at time $t$.

For a treatment-agnostic HyADS interference matrix $\bm{H}_t \in \mathbb{R}^{J \times N}$ at time $t$, let $h_{ijt}$ be the element of $\bm{H}_t$ that denotes the HyADS value representing the influence that power plant $j$ has on ZIP code $i$. For each ZIP code, denote $\mathcal{M}_{it} = \{j: h_{ijt} \ne 0\}$ to be the set of power plants that have a non-zero HyADS value for that ZIP code, and $w_{ijt} = \frac{h_{ijt}}{\sum_{k \in \mathcal{M}_{it}} h_{ikt}}$ to be the normalized weight of power plant $j$ on ZIP code $i$---that is, the proportion of the influence that power plant $j$ has on ZIP code $i$ out of the overall power plant-related pollution that $i$ receives at time $t$. Likewise, let $\mathcal{N}_{jt} = \{i: h_{ijt} \ne 0\}$ be the set of ZIP codes for which the power plant $j$ has a nonzero HyADS value.

Define $Y_{it}$ to be the observed outcome at ZIP code $i$ and time $t$. We propose reconfiguring the outcomes to the intervention unit level via a HyADS-weighted average. Namely, let $Y_{jt} = \sum_{i \in \mathcal{N}_{jt}} w_{ijt}Y_{it}$ be the intervention unit-level outcome for power plant $j$ at time $t$. This quantity is the sum of the outcomes at each of the ZIP codes that is influenced by power plant $j$, weighted by power plant $j$'s relative contribution to the (power plant-related) air pollution in the ZIP code. Such a measure is relevant as it characterizes each power plant's total impact on pollution-related health outcomes. Our analytic framework will be constructed around these power plant-level outcomes. We show in the next section that defining the intervention unit level outcomes in this way leads to an intuitive estimand and identifying assumptions on the intervention unit level, permitting estimation using adaptations of existing methods, but also gives results that have a natural interpretation on the outcome unit level, which may have more direct public health-relevance.

Invoking the potential outcomes framework, let $Y_{jt}(A_{jt},g_j(\bm{A}, \bm{H}_t))$ be the potential outcome for power plant $j$ given its own treatment $A_j$ and spillover value $g_j(\bm{A}, \bm{H}_t)$, some function of all other power plant treatments and the interference network. We discuss a few options and considerations for the mapping $g_j(\bm{A}, \bm{H}_t)$ in Sections~\ref{sec:spilloverspars} and Appendix \ref{sec:networkspars}. Before that, we discuss the target estimand and the causal identifying assumptions required in this setting.

\subsection{Estimands and Identifying Assumptions}\label{sec:assumptions}

Because air pollution is known to have both short-term and long-term effects on health, it is reasonable to hypothesize that the health impacts of an air pollution intervention, such as scrubber installation, might change over time after installation. This type of effect heterogeneity as a function of time since treatment initiation is likely to occur in many applications, and characterizing it may be of substantive interest. Thus, we utilize estimands that are specific to each time point relative to treatment initiation, similar to the approach described above in the dynamic TWFE setting. 

Let $K_{jt}$ denote the number of time periods since the initiation of treatment for unit $j$ at time $t$ (e.g., $-1$ for the year before treatment, 0 for the year of treatment, and so on), with $K_{jt}=-\infty$ if the unit is never treated, and $A_{jt}^k = A_{j}\II(K_{jt}=k)$. Thus, $A_{jt}^k=1$ if the treated unit $j$ is $k$ periods from the initiation of treatment at time $t$ and $A_{jt}^k=0$ otherwise. Then we define the total treatment effect on the treated (TTT) at period $k$ relative to treatment initiation as
\[\tau_{\text{total}}^k = \EE\Big[Y_{jt}\big(1, g_j(\bm{A}, \bm{H}_t)\big) - Y_{jt}\big(0, 0\big) \mid A_{jt}^k = 1\Big]\]
Above, $g_j(\bm{A}, \bm{H}_t)$ in the causal estimand represents a function of the observed value of $\bm{A}$.

In the power plant context, the TTT is the average effect of simultaneously initiating treatment and going from a spillover level of 0 to $g_j(\bm{A}, \bm{H}_t)$, the unit's observed spillover value. In other words, this is the total effect that units experience from their own treatment and the additional spillover effect experienced by treated units. From a policy perspective, regulation of air pollution can be thought of as being composed of two different parts. The first is the regulator--- that is, the authority that sets ambient air quality standards (such as the EPA setting the NAAQS and dictating scrubber use). The second is the local government or power plant operator, who actually enact policies and make decisions on a local level in order to meet the standards set by the higher authority. Here, the TTT is a quantity that would be most relevant to the decision-making of the former--- its value speaks to the overall effect of high-level past policies that led to widespread uptake of scrubber installations in comparison to a business-as-usual world where no such policies were implemented.

Note that $\tau_{\text{total}}^k$ allows for heterogeneity in treatment effects across time, but not treatment initiation cohorts, as discussed in Section~\ref{sec:did_existing}. Our proposed approach will theoretically allow for estimation of effects that are both treatment initiation cohort and time-specific, i.e., $\tau(f,t)$, however we focus on this particular estimand in this work for two reasons. First, in the context of our motivating application, there is less reason to believe that the health impacts of scrubber installations would vary depending on the specific time of installation, particularly given the study time period of 2003-2014. Second, stratifying on both cohort and time simultaneously would result in relatively small sample sizes within each strata.

The following four assumptions are needed for the identifiability of the TTTs. The first three closely follow those of \cite{Butts2021}, while the fourth is a unique condition needed in the bipartite setting.
\begin{assumption}[No anticipation]
Treatment has no causal effect before it is imposed (e.g., there is no ``anticipatory effect'' of the treatment before the treatment period starts for a unit).
\label{assump:anticipation}
\end{assumption}

\begin{assumption}[Treatment is absorbing]
Once a unit is treated, it remains treated. Also, all units are untreated at time $t=1$ (all units begin untreated at the beginning of the study period and treatment does not turn on and off). 
\label{assump:absorbing}
\end{assumption}

\begin{assumption}[Parallel trends (Strict)]
For all units and time periods, the ``unexposed'' potential outcome is given by \[Y_{jt}(0,0) = \mu_j + \lambda_t + \varepsilon_{jt}\]
\label{assump:parallelstrict}
\end{assumption}
This is a strict form of the parallel trends assumption  which imposes a common parallel trend on all units at every time point.

\begin{assumption}[Interference structure is known and treatment-agnostic]
The interference network, defined by the matrix $\bm{H}_t$, is known and is not affected by unit treatment.
\label{assump:agnostic}
\end{assumption}
This assumption is a unique necessity for the exposure-weighted outcome mapping strategy outlined in the previous section, as it ensures that interference weights are not affected by the initiation of treatment over time. For example, consider if the HyADS network introduced in Section~\ref{sec:notation} was affected by scrubber installation, such that the strength of connections between a given power plant and communities decreased by design when a scrubber was installed (indeed this was the case for the standard HyADS used in prior work, which was a direct function of emissions quantities which decrease following scrubber installation). Then, for a power plant that was treated at time $t$, all network edges connected to that plant on or after time $t$ will decrease as compared to prior to time $t$. 
As a result, the power plant-level outcome defined above would also decrease by design, such that outcomes at treated power plants will decline post-intervention even in the absence of any true effects on the health outcome.

An additional property of the TTT introduced above is that it can be reframed as an interpretable quantity in terms of the outcome unit-level treatment effects. To do so, we first define the outcome unit level potential outcome $Y_{it}(\overrightarrow{0})$ to be the outcome that would have been observed in outcome unit $i$ at time $t$ if no intervention units had been treated ($\overrightarrow{0}$ represents a $J$-length vector of zeros corresponding to the treatment status of each intervention unit). 

We make the following additional assumption, defining the untreated and unexposed intervention-level potential outcome $Y_{jt}(0,0)$ in terms of a weighted average of outcome unit potential outcomes under no treatment.

\begin{assumption}[Unexposed potential outcomes] The intervention unit level potential outcome under control is a network-weighted average of the outcome unit potential outcomes under control, i.e.,
\[Y_{jt}(0,0) = \sum_{i=1}^N w_{ijt}Y_{it}(\overrightarrow{0}).\] 
\label{assump:unexposed}
\end{assumption}

This assumption aligns closely with the definition of the intervention unit-level outcomes. Now, note that:
\begin{align*}
    \tau_{\text{total}}^k &= \EE\Big[Y_{jt}\big(1, g_j(\bm{A}, \bm{H}_t)\big) - Y_{jt}\big(0, 0\big) \mid A_{jt}^k = 1\Big]\\
    &= \EE\Big[Y_{jt}\big(1, g_j(\bm{A}, \bm{H}_t)\big) \mid A_{jt}^k = 1\Big] - \EE\Big[Y_{jt}\big(0,0\big) \mid A_{jt}^k = 1\Big]
\end{align*}

Under causal consistency and Assumption~\ref{assump:unexposed}, the following proposition outlines the reframing of the TTT as a function of the outcome unit-level outcomes:

\begin{proposition}
Consider the TTT $\tau_{\text{total}}^k$. Define $\mathcal{Z}^k = \{j: A_{jt}^k = 1\}$ to be the set of power plants for which $A_{jt}^k = 1$.  Then, the sample TTT can be written as \[\widehat{\tau}_{total}^k = \frac{1}{\lvert\mathcal{Z}^k\rvert} \sum_i^N \ell_{it}\big(Y_{it}-Y_{it}(\overrightarrow{0})\big)\]
where $\ell_{it} = \sum_{j \in \mathcal{Z}^k} w_{ijt}$.
\label{prop:ttt}
\end{proposition}

The derivation of this quantity is shown in the appendix.

This reveals that the TTT (when scaled by a factor of $\lvert\mathcal{Z}_t^k\rvert/\sum_i \ell_{it}$) can be interpreted as a weighted average of outcome unit level treatment effects $Y_{it}-Y_{it}(\overrightarrow{0})$, with outcome units weighted based on the relative degree to which they are connected with the units that are $k$ periods from treatment. Effectively, the outcome units that are heavily influenced by the power plants $k$ periods from treatment are up-weighted in this estimand and those that are not are down-weighted. This representation allows for the TTT to be defined on the more intuitive intervention unit level, but additionally be interpreted on the outcome unit level, adding to its relevance from a policymaking perspective. 

\subsection{Estimation}\label{sec:estimation}

In this section, we outline a DiD strategy following the approaches of \cite{Butts2021} and \cite{Gardner_twoway} for the estimation of staggered treatment effects while accounting for spillover effects and discuss some considerations when adapting their approach to the BNI setting. Identification of the TTT is shown in the appendix.

Estimation is proposed as a two-stage DiD estimator:
\begin{enumerate}
    \item Estimate $Y_{jt} = \mu_j+\lambda_t+\varepsilon_{jt}$ for control units (units that either never experienced treatment or were not-yet-treated, and were also unexposed to spillover), and obtain estimates for fixed effects $\hat{\mu}_j$ and common trend $\hat{\lambda}_t$.
    \item Use the model to ``predict'' outcomes under control for \textit{all} units and time points, i.e., $\widehat{Y}_{jt}(0,0)=\widehat{\mu}_j+\widehat{\lambda}_t, \ \ \forall (j,t)$. Use these predictions to compute $\tilde{Y}_{jt}=Y_{jt}-\widehat{Y}_{jt}(0,0)$, i.e., the difference in the observed outcome and the predicted outcome under control.
    \item Regress $\tilde{Y}_{jt}$ for all units and time points on dummy variables $A_{jt}^k$. The corresponding regression coefficients are the $\widehat{\tau}^k_{total}$.
\end{enumerate}

In the above procedure, the $\tilde{Y}_{jt}$ values estimate individual effects $\tau_{jt}$. Therefore, regressing the individual effect estimates on treatment/exposure dummy variables $A_{jt}^k$ provides average effect estimates for groups of treated units at particular lead/lag periods $k$. \cite{Gardner_twoway} shows that $\mu_j$ and $\lambda_t$ in the above procedure are identified from the subpopulation of untreated groups and periods as long as untreated and treated observations exist for each group and period. Unbiasedness, consistency, and asymptotics (since $Y_{jt}$ is obtained from estimates from the first stage of the above procedure) are shown and discussed in \cite{Gardner_twoway}. Estimation in steps 1 and 3 above can be done using ordinary least squares.

The key to adopting this approach (and in most other DiD approaches as well) is that comparison groups (that is, the groups of never-treated or not-yet-treated) need to be defined.  In the standard non-bipartite, no-interference case, this is a relatively simple matter, since treatments are well-defined and limited to the treatment status of each unit individually. In the non-bipartite case with interference, control units would be defined as units that were untreated and ``unexposed'' (i.e., no spillover exposure). This spillover exposure value is defined by the choice of $g_j(\bm{A}, \bm{H}_t)$, which maps the vector of all treatments (except for a unit's own treatment) to some value. In order to minimize information loss, for example, one might consider a unit's spillover exposure value defined by a form of weighted sum of other units' treatments. This definition would retain the full amount of treatment data, as opposed to simpler methods. With a relatively dense interference network, as is the case in many real-world applications, using this mapping, it would likely be quite difficult to find units that naturally have zero spillover exposure. In the BNI setting, this idea becomes more complex, though similar logic applies. At the intervention level, untreated units are easy to identify. However, even defining spillover at the intervention unit level is complex and units with true zero spillover values may not exist in many definitions and cases.

In our motivating application, the exposure-response matrix of power plant emissions to health outcomes is very dense. This means that there is unlikely to be a (sufficiently large) set of clearly unexposed intervention units. However, despite the density of the network, in this instance the actual network edge weights (e.g., effect of the power plant $\mathcal{A}$'s emissions on ZIP code $\mathcal{X}$) are highly right skewed, meaning there are a large number of edges with low weights and a relatively small number of edges with large weights. As a result, one option to ensure the existence of enough ``unexposed'' units is to induce sparsity in the network under the assumption that some edge weights may be sufficiently small such that they could be thresholded to zero. In practice, this could be achieved in a couple of different ways. First, one could directly induce sparsity in the network itself---i.e., directly set some values in $\mathbf{H}$ equal to zero, and then compute spillover values (which we refer to as ``network sparsification''). Otherwise, another option is to first compute the desired measure of spillover for each unit, and then induce sparsity by setting some values equal to zero (or ``spillover value sparsification''). The spillover value sparsification approach is detailed below, while the network sparsification approach is discussed in the appendix.

\subsubsection{Spillover Value Sparsification}\label{sec:spilloverspars}

A method to guarantee sparsification is to take advantage of the known interference network and employ a mapping approach to define what a spillover effect is in this context. This idea differs from network sparsification in that rather than sparsifying the interference matrix and then defining spillover, the spillover effect is first defined, with thresholding to induce sparsification done afterwards.

To define $g_j(\mathbf{A}, \bm{H}_t)$ in the context of a dense bipartite network, we propose the following approach. The high level idea is that, to get the spillover for an intervention unit $j'$, we (1) for each outcome unit compute a measure representing the `degree of treatment' that unit experiences from power plants besides $j'$, denoted by $m_{ij't}$ and (2) map these outcome unit level values back to intervention unit $j'$ using a network connection-weighted average to get $g_{j'}(\mathbf{A}, \bm{H}_t)$. Thus, $g_{j'}(\mathbf{A}, \bm{H}_t)$ roughly measures the degree to which the outcome units most affected by intervention unit $j'$ are influenced by other treated units. Formally, $g_{j'}(\mathbf{A}, \bm{H}_t)$ is computed using the following procedure:
\begin{enumerate}
    \item For each outcome unit $i$ and time point $t$:
    \begin{enumerate}
        \item Let $h_{ijt}A_{jt}$ be a weighted treatment value for outcome unit $i$ from intervention unit $j$ (i.e., interference matrix weight multiplied by treatment).
        \item Define the continuous treatment value $m_{ij't}$ by all intervention units except for $j'$ as \[m_{ij't} = \sum_{j \ne j'} h_{ijt}A_{jt}.\]
    \end{enumerate}
    \item Compute a spillover treatment value for $j'$, defined as a weighted sum of the $m_{ij't}$ values weighted by $h_{ij't}$: \[g_{j'}(\bm{A}, \bm{H}_t) = \sum_{i=1}^n h_{ij't}m_{ij't}\]
\end{enumerate}

\begin{figure}[ht!]
    \centering
    \includegraphics[width=\textwidth]{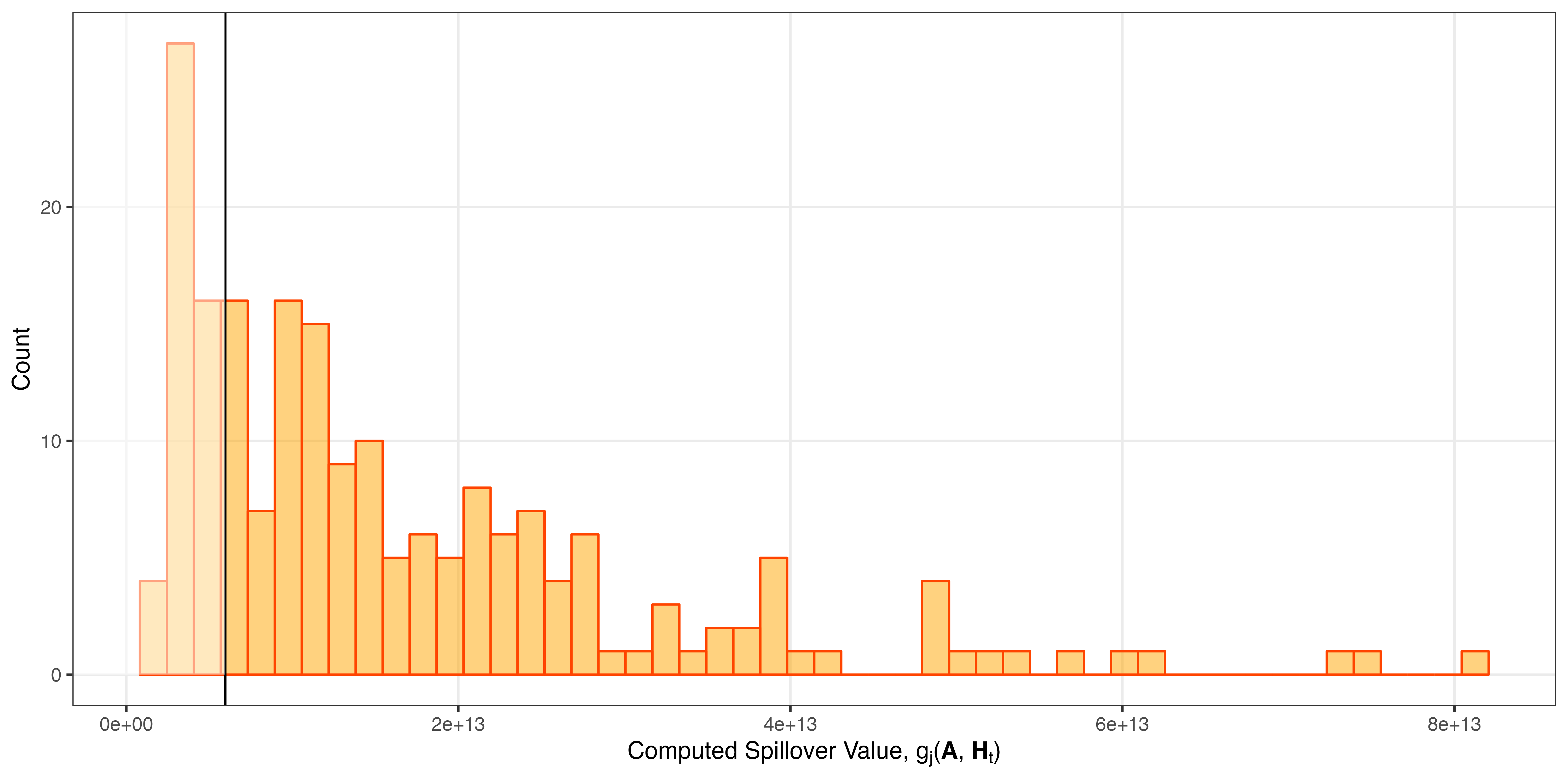}
    \caption{Histogram of computed spillover values $g_{j}(\bm{A}, \bm{H}_t)$ at power plant locations, using the spillover value sparsification approach based on scrubber installation status on power plants in 2005. The vertical black line denotes the 25$^{\text{th}}$ percentile value and the shaded area to the left shows the units whose spillover values would be thresholded using this approach.}
    \label{fig:spillover_hist}
\end{figure}

\begin{figure}[ht!]
    \centering
    \includegraphics[width=\textwidth]{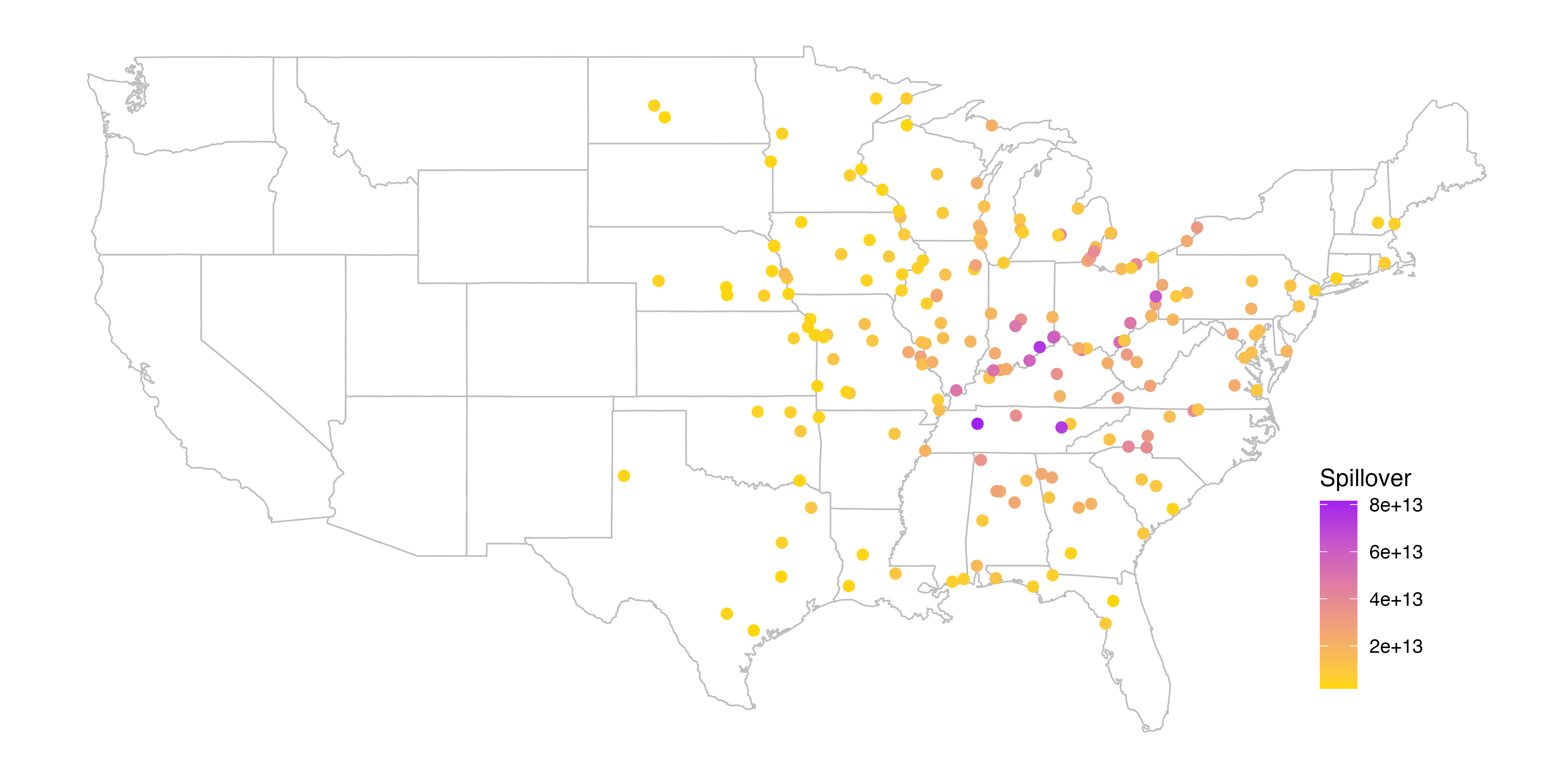}
    \caption{Map of computed spillover values $g_{j}(\bm{A}, \bm{H}_t)$ at power plant locations, using the spillover value sparsification approach based on scrubber installation status on power plants in 2005.}
    \label{fig:spillover_map}
\end{figure}

Computing the spillover treatment value $g_{j}(\bm{A}, \bm{H}_t)$ for each intervention unit $j$, we can then apply a threshold, setting all computed spillover values beneath this threshold to zero. As a result, intervention units that are untreated themselves, in addition to having a spillover value under the threshold, are considered controls. In the context of our motivating application, an example of spillover values computed using this approach in a single particular year is shown in Figures~\ref{fig:spillover_hist} and \ref{fig:spillover_map}.

\section{Case Study: Effects of Power Plant Scrubber Installations on Coronary Heart Disease Hospitalizations From 2003-2014}\label{sec:case}

In this section, we discuss the application of our proposed approach to the case study introduced in Section~\ref{sec:motivating}. We studied the effect of SO$_2$ FGD scrubber installations during a 12-year period from 2003 to 2014 on the rates of coronary heart disease hospitalization among older Americans in the contiguous U.S. Power plant facilities that had SO$_2$ scrubbers installed in 2003 or prior (per Assumption 3.2) and those that were not in continuous operation during the entire study period were excluded from the analysis. Furthermore, power plants located on the West Coast and Mountain West states in addition to Alaska and Hawaii were excluded due to a relative lack of coal-fired power plant activity in these regions. Treatment spillover values were computed using the spillover value sparsification approach outlined in Section~\ref{sec:spilloverspars} in each year for each of the remaining 197 power plants. A threshold at the 25$^{\text{th}}$ percentile was applied to the spillover values at each year. The resulting treatment and spillover indicators are visualized in Figures~\ref{fig:treatment} and \ref{fig:spillover}. At the ZIP code level, the outcomes are defined as annualized rates of CHD hospitalizations per 10,000 person-years in Medicare fee-for-services beneficiaries. Power plant-level outcomes are computed as described in Section~\ref{sec:notation}.

To account for potential time-varying confounding, meteorological variables (relative humidity, temperature, and precipitation totals) were obtained from the University of Idaho Gridded Surface Meteorological Dataset (GRIDMET) and spatially averaged to ZIP codes \citep{Abatzoglou_gridmet_2011}. Data on counties that were found to be in nonattainment with the 1997 and 2006 NAAQS were obtained from the EPA Green Book of Nonattainment Areas for Criteria Pollutants \citep{usepa_greenbook_2024}. This data was translated to the ZIP code-level for analysis. Although the announcement of the 2012 NAAQS occurred prior to the end of the study period, nonattainment areas were not designated until 2015 (after the study period), and therefore were not relevant for this analysis. NAAQS nonattainment designation was identified as surrogate measure of whether localities potentially instituted alternative SO$_2$ reduction in response to impending NAAQS standards implementation. Therefore, this variable was included as a potential confounder. ZIP-level confounders were translated to the power plant level via the same approach applied to the outcomes as in Section~\ref{sec:notation}. 

The estimation of the TTT was done according to the procedure outlined in Section~\ref{sec:estimation}, using the did2s R package \citep{did2s}. For each year, power plants that were both untreated and had spillover values under the 25$^{\text{th}}$ percentile threshold computed above were used to estimate the unexposed outcome $Y_{jt}(0,0)$. After residualizing to obtain the treatment effect estimates of $\tau_{jt}$ for all observations, these were regressed on dummy indicator variables for relative year since scrubber installation $A_{jt}^k$.

\begin{figure}[ht!]
    \centering
    \includegraphics[width=0.8\textwidth]{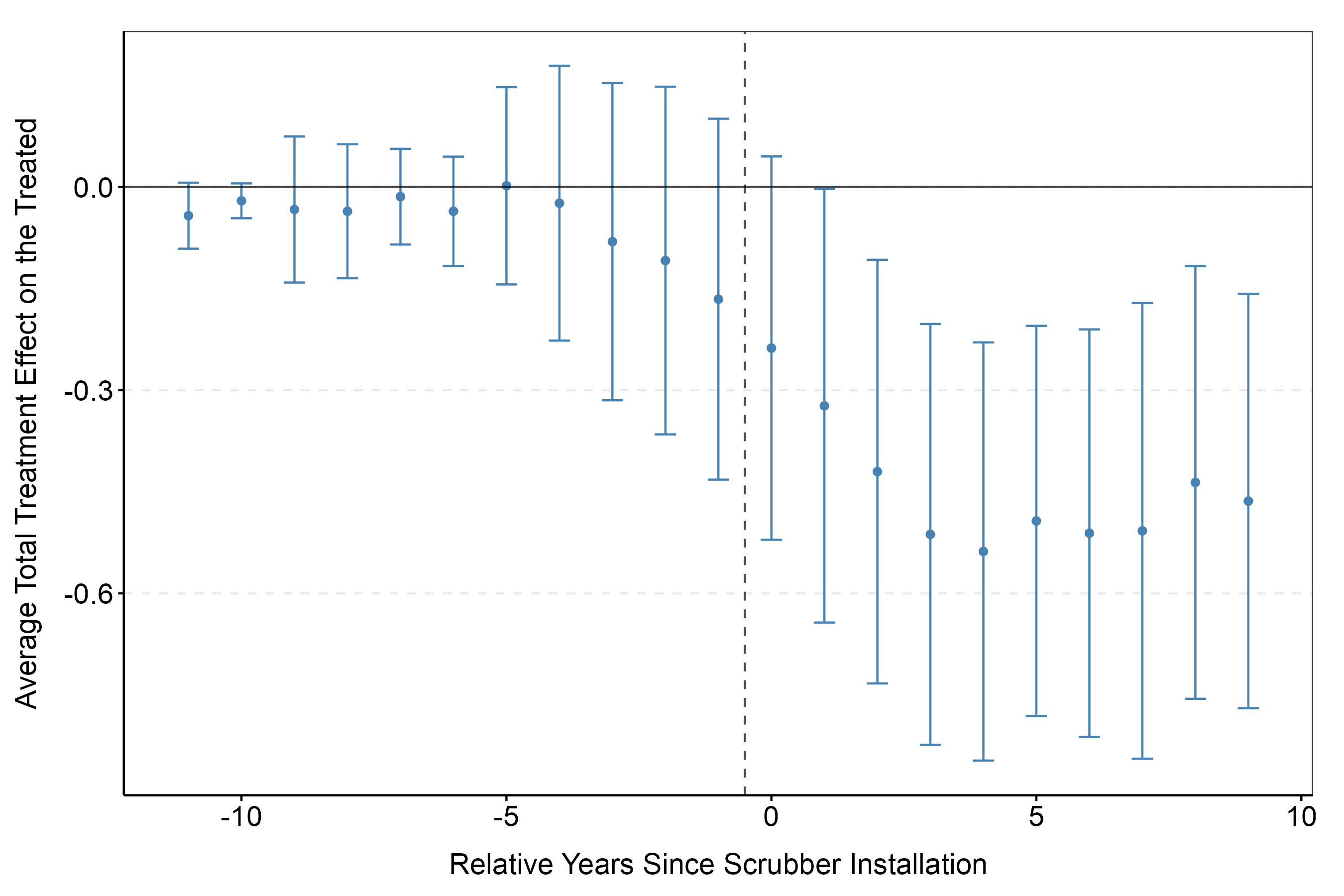}
    \caption{Estimated average total treatment effect on scrubber-installed plants on coronary heart disease hospitalization rates, 2003-2014.}
    \label{fig:results}
\end{figure}

The full results are presented in Figure~\ref{fig:results}. Negative values for relative years since scrubber installation indicate pre-treatment periods, with time 0 being the year in which scrubbers were installed. No significant pre-treatment values differed significantly from zero, indicating no deviation from the parallel trends assumption \citep{Gardner_twoway}. In the post-treatment lag periods, we observe statistically significant negative values for the TTT, indicating decreases in CHD hospitalization rates. The effects appear to reach their peak magnitude approximately 3 years post-scrubber installation and remain fairly constant in later periods. This apparent lag in effects may have a number of explanations. First, longer-term health effects of emissions reductions may take time to materialize in affected populations. Second, in our analysis we define a power plant facility as ``treated'' during the year in which scrubbers were first installed. However, larger facilities may have longer ramp-up times during which scrubbers are installed at multiple EGUs, leading to a delay in discernible effects. Overall, these results suggest that the installation of SO$_2$ FGD scrubbers during this time period had a protective effect against emissions-related CHD hospitalizations. To interpret these estimates on the outcome unit level, we apply the rescaling procedure outlined in Section~\ref{sec:assumptions} on the estimated TTT. For example, at 3 years post-scrubber installation, the estimated weighted average effect of scrubber installations is -22.5 (95\% confidence interval:[-40.1, -4.9]), meaning an average decrease of 22.5 CHD hospitalizations per 10,000 person-years on the ZIP code level.

\section{Conclusion}\label{sec:conclusion}

In this paper, we have considered causal analyses for multi-group, multi-period panel data in the context where units on which treatments are imposed are disjoint from units on which outcomes are measured. In this setting, these intervention units are connected to the outcome units in a dense bipartite network, creating several obstacles when the goal of a study is to estimate treatment effects over time. We discuss the existing difference-in-differences methodology and the particular challenges that arise when attempting to extend such methodology to studies in the BNI setting. These considerations are made in the context of a case study of SO$_2$ emissions reduction technology installation on coal-fired power plants and their effects on coronary heart disease hospitalizations among older Americans.

To contribute to the growing literature on difference-in-differences, particularly in the presence of treatment interference, we introduce a general framework for quasi-experimental analyses under complex bipartite network interference structures. We propose a data reconfiguration approach in which we take advantage of the information provided by a known interference network to map the outcome-level data to the intervention level. We also provide a rescaling approach to interpret these results on the more policy-relevant outcome level. Through this approach, we leverage the existing DiD methodology to estimate treatment effects and find that SO$_2$ scrubber installations in the 2000s and early 2010s had an overall beneficial effect in reducing CHD hospitalization rates in the Medicare population. 

Our work has several limitations. The complexity of the BNI setting necessitated the mapping of values between the two different unit types, complicating the methodology and interpretations. However, the feasibility of DiD-type analyses in this setting is a nontrivial methodological challenge, and relies on such approaches. The sparsification procedures discussed in this paper also represent simplifications of the full interference network; in practice, one may consider data-driven approaches to more accurately preserve the underlying interference structure. The identifying assumptions of no anticipation and parallel trends made in this paper are rather stringent. However, extensions to allow for limited anticipation and weaker parallel trends assumptions are subjects of ongoing work. Similar relaxations of such assumptions are discussed in several recent works \citep[see][]{Callaway2021,Butts2021,Rambachan_parallel_2023}. 

This work is, to our knowledge, the first to examine an extension of quasi-experimental event study methodology to consider bipartite network interference. Likewise, it is an important contribution to the fast-expanding literature on the causal analysis of policy effects in the bipartite setting by examining effects over time, rather than at a single time point. The methodology from this perspective is increasingly relevant where simpler estimation strategies fail in these complex scenarios. From an environmental policy standpoint, such analyses can provide a more comprehensive view of how regulations such as the NAAQS have affected human health in the past and inform the enactment of future policies.

\clearpage
\bibliographystyle{apalike}
\bibliography{references}

\clearpage
\section*{Appendix}
\renewcommand{\thesubsection}{\Alph{subsection}}
\renewcommand\thefigure{\thesubsection.\arabic{figure}}   
\renewcommand\thetable{\thesubsection.\arabic{table}}   
\setcounter{page}{1}

\subsection{Proofs}

\subsubsection{Identification of the TTT.}
Under Assumptions~\ref{assump:anticipation}-\ref{assump:agnostic}, the identification of the TTT can be argued as follows, as in Appendix A of \cite{Butts2021}:
\begin{align*}
    \EE\Big[Y_{jt}\big(1, g_j(\bm{A}, \bm{H}_t)\big) - \mu_j - \lambda_t \mid A_{jt}^k = 1\Big] &= \EE\Big[Y_{jt}\big(1, g_j(\bm{A}, \bm{H}_t)\big) - Y_{jt}(0,0) + \varepsilon_{jt} \mid A_{jt}^k = 1\Big]\\
    &= \EE\Big[Y_{jt}\big(1, g_j(\bm{A}, \bm{H}_t)\big) - Y_{jt}(0,0) \mid A_{jt}^k = 1\Big]\\
    &= \tau_{\text{total}}^k
\end{align*}

\subsubsection{Proof of Proposition~\ref{prop:ttt}.}

Defining $\mathcal{Z}^k = \{j: A_{jt}^k = 1\}$ to be the set of power plants for which $A_{jt}^k = 1$, and $\ell_{it} = \sum_{j \in \mathcal{Z}^k} w_{ijt}$, 
\begin{align*}
    \widehat{\tau}_{total}^k &= \frac{1}{\lvert\mathcal{Z}^k\rvert} \sum_{j \in \mathcal{Z}^k} Y_{jt} - \frac{1}{\lvert\mathcal{Z}^k\rvert} \sum_{j \in \mathcal{Z}^k} Y_{jt}(0,0)\\
    &= \frac{1}{\lvert\mathcal{Z}^k\rvert} \sum_{j \in \mathcal{Z}^k} \sum_i^N w_{ijt}Y_{it} - \frac{1}{\lvert\mathcal{Z}^k\rvert} \sum_{j \in \mathcal{Z}^k} \sum_i^N w_{ijt}Y_{it}(\overrightarrow{0})\\
    &= \frac{1}{\lvert\mathcal{Z}^k\rvert} \sum_i^N Y_{it} \sum_{j \in \mathcal{Z}^k} w_{ijt} - \frac{1}{\lvert\mathcal{Z}^k\rvert} \sum_i^N Y_{it}(\overrightarrow{0}) \sum_{j \in \mathcal{Z}^k} w_{ijt}\\
    &= \frac{1}{\lvert\mathcal{Z}^k\rvert} \sum_i^N \ell_{it} Y_{it} - \frac{1}{\lvert\mathcal{Z}^k\rvert} \sum_i^N \ell_{it} Y_{it}(\overrightarrow{0})\\
    &= \frac{1}{\lvert\mathcal{Z}^k\rvert} \sum_i^N \ell_{it}\big(Y_{it}-Y_{it}(\overrightarrow{0})\big)
\end{align*}
The first equality follows from causal consistency and the second follows from the definition of the intervention unit-level outcomes and Assumption~\ref{assump:unexposed}.

\subsection{Network Sparsification}\label{sec:networkspars}
\setcounter{figure}{0}

The network sparsification approach entails, as the name suggests, directly inducing sparsity on the interference network defined by the matrix $\bm{H}_t$. In this context, we first discuss the motivation and implications of  sparsification on the network. On a bipartite network, an untreated and unexposed control unit is an intervention unit that does not receive any effects of treatment on the network. In practice, this would be, for example, untreated power plants for which all connected ZIP codes are only connected to other untreated power plants. This ensures that the computed outcome measure associated with each untreated plant is truly unexposed to treatment. Under this definition, only first-level mutual connections matter-- that is, an untreated power plant is considered sufficiently ``unexposed'' if its direct ZIP code connections are not themselves connected to any treated power plants, but any connections past this (e.g., to other ZIP codes that are then connected to treated plants) are not relevant. 

\begin{figure}[ht!]
    \centering
    \includegraphics[width=\textwidth]{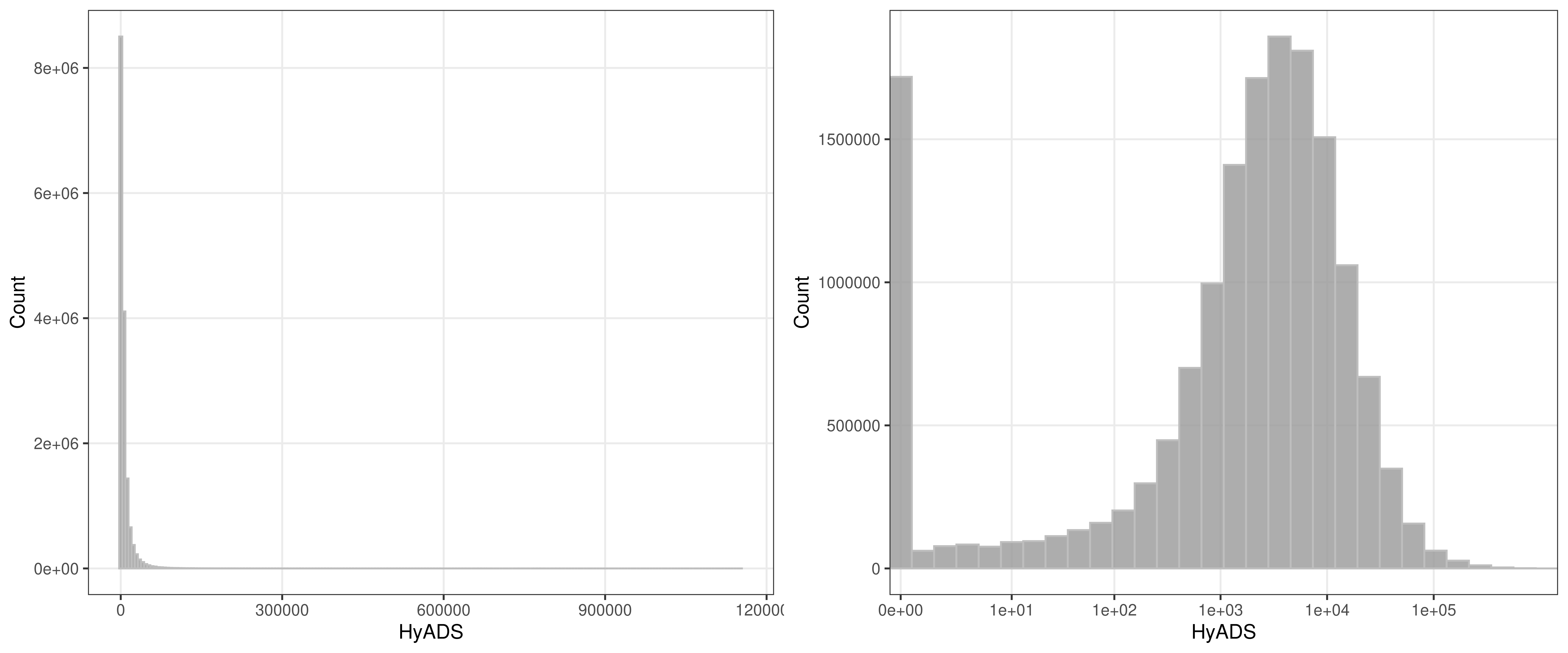}
    \caption{Histogram of HyADS values over entire network in 2005 (before any sparsification), with the same plot on the log$_{10}$ scale on the right.}
    \label{fig:hyads_unspars}
\end{figure}

Network sparsification can be conducted in a relatively simple way, such as imposing an absolute threshold on network edges, or in a data-driven manner, such as the statistically validated network approach of \cite{tumminello_statistically_2011} which is applied to bipartite networks in \cite{Bongiorno_validated_2017}. Post-sparsification, one would still need to apply a mapping approach to obtain a spillover treatment value $g_j(\bm{A}, \bm{H}_t)$ for each unit. The downside of the network sparsification approach is that it may not result in sufficient sparsification of the spillover values to allow for DiD analysis, depending on the structure of the network and the spillover mapping approach. This is particularly true for very dense networks and may require further imposing restrictive assumptions or clustering to achieve enough unexposed units. Within our motivating application, the transportation of air pollution through meteorological and atmospheric processes means that there are very few HyADS values that are naturally zero-valued. As an example, Figure~\ref{fig:hyads_unspars} shows the HyADS values observed in 2005. For a plant to be completely unexposed to spillover, all ZIP codes that are connected to the plant would have to \textit{only} be connected to untreated plants. In practice, this is extremely unlikely to occur for enough units to provide sufficient controls for DiD analysis.

\subsection{Additional Figures}
\setcounter{figure}{0}
\begin{figure}[ht!]
    \centering
    \includegraphics[width=\textwidth]{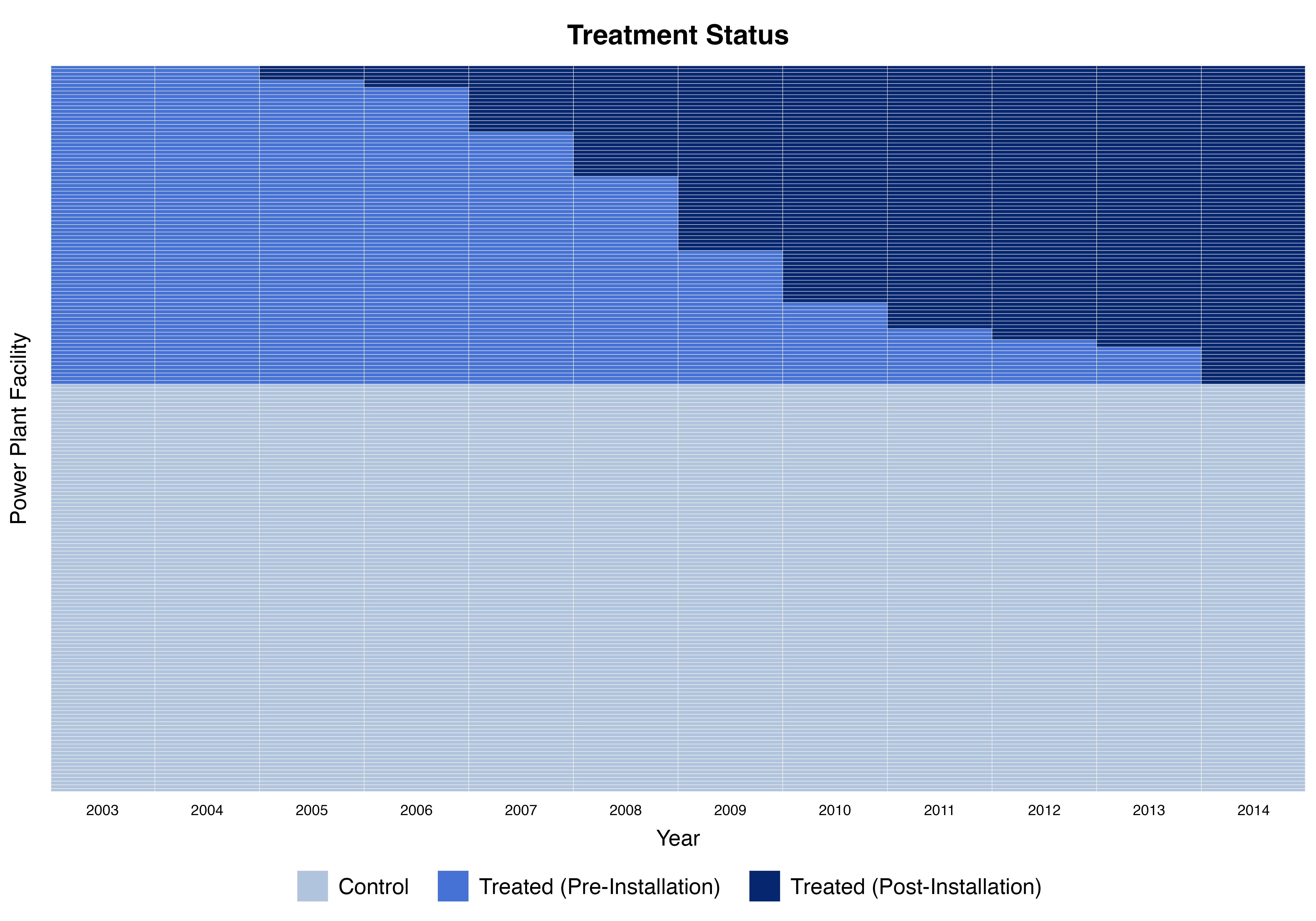}
    \caption{Scrubber installation status on power plants, 2003-2014.}
    \label{fig:treatment}
\end{figure}

\begin{figure}[ht!]
    \centering
    \includegraphics[width=\textwidth]{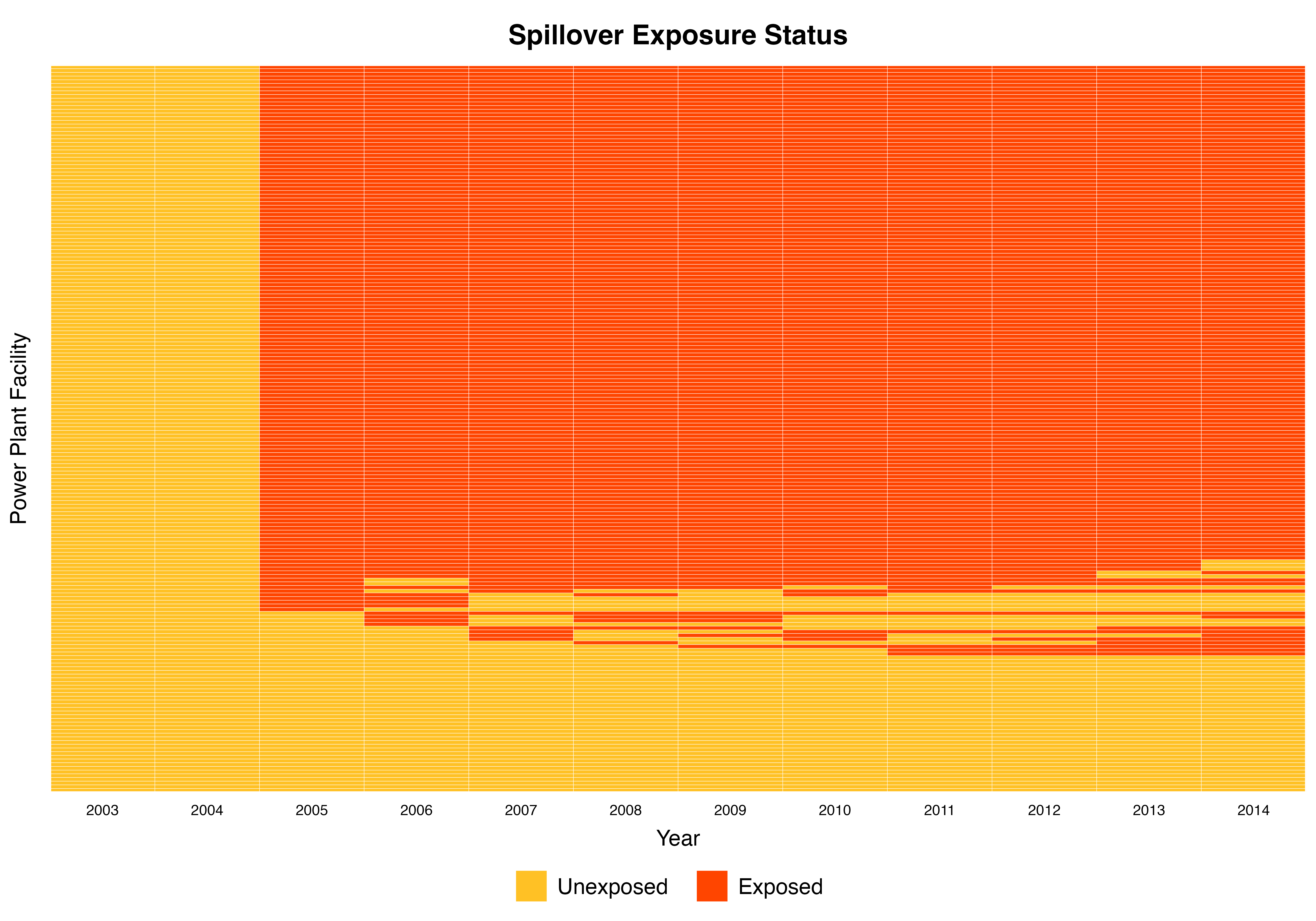}
    \caption{Spillover exposure status on power plants, thresholding at the 25th percentile, 2003-2014.}
    \label{fig:spillover}
\end{figure}

\end{document}